\documentclass[aps,prl,showpacs,showkeys,twocolumn]{revtex4}

\newcommand{\beq}{\begin{equation}}
\newcommand{\eeq}{\end{equation}}
\newcommand{\bef}{\begin{figure}}
\newcommand{\eef}{\end{figure}}

\usepackage{graphicx}   % need for figures
\usepackage{amssymb}
\usepackage{pifont}
\begin{document}

\title{Growing E-coli in the presence of electric fields}
\author{Eshel Faraggi}
\email{efaraggi@gmail.com}
\affiliation{Research and Information Systems, LLC, 1620 E. 72nd ST., Indianapolis, IN 46240, USA}
%\date{\today}

\begin{abstract}
Experiments were performed to test the effect of an electric field on growing E-coli cultures. Cell growing dishes were fitted with platinum wires to create a potential drop across the dish. Water diluted E-coli cultures were brushed across the dish in various patterns and the growth of E-coli colonies was recorded. Repeated experiments on this model were performed. It was found that E-coli grew preferentially in the vicinity of the lower electric potential, a region characterized by the presence of positive ions while there was a clear non-growth area near the higher potential, a region characterized by the presence of negative ions. These results support previous theoretical analysis of the general problem of cell division which accounts for the symmetry of this system, and in which electrostatic repulsion is the primary long range driving force of the dividing biological cell.

\end{abstract}


\maketitle

%\section{Introduction}
%
Biological cell division is the cornerstone of life.~\cite{preston1990increased,lutkenhaus1997bacterial,rothfield1999bacterial,croce2005mirnas,faraggi2010electrostatic} For all organisms, production of viable progeny depends on the successful replication of
DNA and separation of these replicas to new copies. Multi-cellular life result from many consecutive cell divisions, single cell organisms divide to form copies of themselves, while viruses use cell division in other organisms to produce copies of themselves. From a practical stand point, a deeper understanding of cell division and the ability to better manipulate it will revolutionize medicine. Since cell division is essential for life, required by the first primordial cell/s, the fundamental underlying interactions responsible for it must be general, and most likely involve only simple direct interactions that can result by chance in the original sea of life. However, countless generations of mutations from these original cells masked the underlying interactions responsible for cell division with complex processes. Prokaryotic cells probably offer a closer version of the division of primordial cells. Here, wildtype Escherichia-coli (E-coli) is the model organism chosen for investigation. 

All life shares DNA as the basic code of its particular form. All life also shares the twenty amino-acids that make up proteins, the matter of life. This points in the direction of a common ancestor. It is also plausible to assume that the mechanisms for segregation of DNA copies will rest on the same fundamental physics across different life forms. There are four known fundamental forces. Two of them, the strong and weak nuclear forces, have to do with interactions on the atom's nucleus length scale. Per current knowledge, these forces do not contribute to life beyond their role in the creation of matter. A third force, gravity, has to do with large length scales, it is negligible in the biochemistry of living organism. This leaves us with the electromagnetic force as the sole interaction responsible for the differences between living and non-living matter. In fact, quantum electromagnetism is the basis of atomic structure and interaction. When moving from a few atoms to the many atom systems prevalent in living organisms it is found that quantum effects become mostly negligible, and classical electromagnetism (EM) is used. EM is used throughout biology.~\cite{ryaby1978modification,luben1991effects,honig1995classical,kirson2004disruption,bassett1991long,griffin2008role}

%Here a set of experiments are described. In these experiments the effect of an electric field on growing E-coli cultures was tested. After a brief review, including of a proposed model for the mechanics of cell division, the experimental model is described, followed by results, and discussion. 

Current understanding of E-coli, and bacterial cell division in general, relay on several key proteins. Some proteins were identified for the process of separating the genetic material such as the F-plasmid-partitioning proteins SopAB~\cite{ogura1983partition,lim2005bacterial} and the Walker-type ATPases ParAB.~\cite{ebersbach2001double,barilla2005bacterial,fogel2006dynamic,pratto2008streptococcus,ptacin2010spindle,hwang2013mediated,vecchiarelli2013cell} While the tubulin-like FtsZ protein was identified as forming a ring structure mid cell and facilitating the final stages of separation into two cells.~\cite{errington2003cytokinesis,adams2009bacterial,erickson2010ftsz,de2010advances,lutkenhaus2012bacterial,egan2013physiology} Control over the function of FtsZ is accomplished by the three proteins MinCDE which bind selectively to the cell membrane guiding the location and activity of FtsZ.~\cite{leonard2005towards} 

There is no doubt these are important factors in the division of E-coli and other types of evolved bacteria we find today. However, cell division was required by primordial, original cells. Four main criticisms of this overall picture of cell division of the most rudimentary cells come up: 1) These processes require specific interactions between multiple specific macromolecules that were developed over many generation of evolution. These mechanisms are too complex to occur by chance in primordial cell division. 2) These models lack a long-range component and are localized to contact forces between proteins. It is unclear how such forces localized to nanometer distance can give rise to the directed micrometer movements associated with cell division. 3) No account for the directed force required to elongate the cell, separate the DNA, etc. Account is only given to available energy in the form of ATP to complete the division, this energy in this case in non-directed however. 4) No account for the separation and opposite motion of the two volumes delineating the daughter cell area in the original cell. These areas are mostly composed of water~\cite{milo2009bionumbers}, and undergo directed motion on the cell length scale. 

We see then that much is left to be desired in our current understanding of cell division. Especially for the mechanics of separation at its most basic level. A level appropriate for those first instances of cell division that occurred in nature. For organism that had no time to mutate, or even create, their proteins to preform complicated tasks.

Previously, a theoretical framework that accounts for the gross features of cell division was introduced by the present author.~\cite{faraggi2010electrostatic,faraggi2012symmetrical} An outline follows, its main feature is showing that a symmetric two charge configuration in an ionic solution experiences a heightened interaction due to the rearrangement of the charged ions. While the individual electromagnetic interaction in ionic solution is exponentially attenuated as prescribed by Debye screening, the many-body collective force of repulsion between the two symmetrical parts is increased as compared to charges in non-ionic media.

As is well known,~\cite{jackson} the potential $\Phi$ for a point charge, $q_1$, in an ionic solution exponentially decays with a length scale given by the Debye length. Since the Debye length for biological systems is of the order of a few protein helix turns it was judged that electrostatic repulsion is of negligible importance in cell division. However, the dividing biological cell is inherently a symmetric process. This implies that we are interested in the interaction between two charges. The modified Poisson equation for two identical charges, $q_1 = q_2 = q$, a distance of $2 x_0$ apart, and immersed in an ionic solution with dielectric constant $D$, is given by~\cite{faraggi2012symmetrical}
\beq
\scriptstyle{D \cdot \nabla^2 \Phi = - 4 \pi q [ \delta ({\mathbf x}+{\mathbf x}_0) + \delta ({\mathbf x}-{\mathbf x}_0)] - 4 \pi q_m n_0 [ e^{\frac{-q_m \Phi}{k_B T}} - 1 ],}
\label{debyeq2}
\eeq
where ${\mathbf x}$ is a vector position, $n_0$ the particle density, $q_m$ the ions charge, $k_B T$ the thermal energy, and $\delta(x)$ the Dirac delta function.

\begin{figure}[th!]
\includegraphics[scale=0.7]{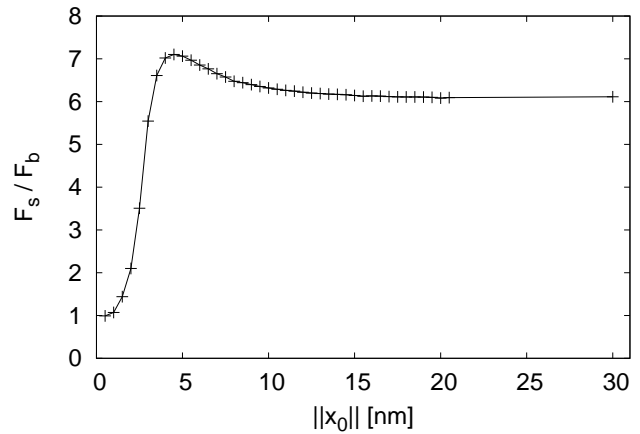}
\caption{The ratio of the electrostatic force between two half ionic volumes with identical charges to the same configuration in a dielectric. This ratio is plotted as a function of the separation distance between the charges. Additional interactions from ions reacting to the charges introduce the added repulsion we see in the plot. From previous work.~\cite{faraggi2012symmetrical}}
\label{res1}
\end{figure}

Integrating the solution of Eq.~\ref{debyeq2} we can calculate the force of separation between the two daughter cells represented by the symmetrical half-volumes containing $q_1$ and $q_2$, i.e., the force of separation between the charge densities of the two volumes delineated by the symmetry of this problem.~\cite{faraggi2012symmetrical} Note that ions in different half volumes interact with each other and contribute to the overall repulsion between the two half volumes.  On the y-axis of Fig.~\ref{res1} we have the ratio of the separating force between two charges in an ionic solution to the Coulomb force without an ionic solution. On the x-axis of Fig.~\ref{res1} we have the distance between the charges $q_1$ and $q_2$. In this case both $q_1$ and $q_2$ equal the charge of a single proton. We see that the ratio reaches a plateau of about 6 at around 4 nm. That is, due to the symmetry between the charge distribution in the two half volumes, the repulsive force between the two half volumes due to $q_1$ and $q_2$ is increased six-fold as compared to their Coulomb repulsion. This ratio remains approximately constant over the range of the calculation. As was shown previously, this ratio is significantly increased if the amount of charge in $q_1$ and $q_2$ is increased.~\cite{faraggi2012symmetrical}

To test these ideas the following experimental system was set up. The set up starts with a clear disposable Petri dish (Fisherbrand 100$\times$15mm). In this dish two platinum electrodes (0.009~in diameter) were fashioned, centered in the Petri dish at a distance of 5~cm apart, and held in place by small pieces of clear packaging tape. The Petri dish was then filled with 25~ml Agar growing medium. This amount of Agar was found optimal for the dish size used. The Agar was prepared using the following protocol: 10~g Difco Agar, 10~g NaCl, 10~g Tryptone, and 5~g Yeast Extract, were combined dry then poured over 1~L of distilled water, mixed, and cooked in the a microwave until completely homogeneous. For each experiment, after fitting them with electrodes, Petri dishes were filled with 25~ml of this growth medium, and set aside to cool down.

After cooling, E-coli was seeded on the surface by gently brushing a cotton swab impregnated with E-coli solution on the Agar surface in a zigzag pattern. E-coli solution was prepared by depositing three swabs from grown E-coli colonies into approximately 50~ml of distilled water. The E-coli solution was shaken for 10 seconds using a radial shaker before each use.  Two zigzag patterns were brushed on a single dish, one over each electrode. A center line of E-coli was brushed in some cases. Electrodes were then connected to a DC power supply (TE HY1803DL) arbitrarily selecting which electrode to connect to which pole of the power supply. The dishes were then deposited in an incubator (ESPEC BNA-211) held at 38$\pm 0.5^{\circ} C$. Before acquiring this incubator, an electric heat-pad in a regular two foot cooler served as an incubator at 36$\pm 2^{\circ} C$. Experiments carried in it produced similar results.

\begin{figure}[th!]
\includegraphics[scale=0.3]{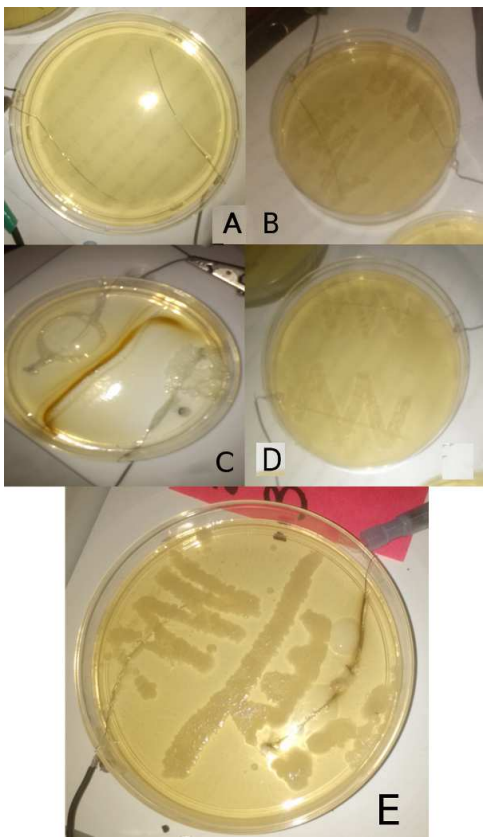}
\caption{Preparation and representative results. Initial system (A), control experiment with no potential difference between electrodes (B), a dish after 24 hours at 5~V (C), a dish after 24~h at 1.5~V (D), a dish after 24~h at 3~V. For all 3~V cases only one zigzag pattern grew completely, and selectively around the lower potential electrode. This seems to indicate EM fields play a crucial role in the process of biological cell division.}
\label{dish.init}
\end{figure}

Fig.~\ref{dish.init}A gives the image of the system immediately after the E-coli was deposited and the electrodes connected to the power supply. Note that no visible zigzag pattern appear due to the low concentration of E-coli cells, only after the growth of colonies (approximately 8 hours) the zigzag pattern appears complete. Direct Current (DC) potentials of 1.5, 3, and 5~V were tested in separate experiments. Over the course of several months, experiments were performed on over 50 dishes at 3~V, and about 10 dishes for each of the other cases. 
The observable was if a continuous zigzag patterns grew, and over which electrode. Due to initial size constraints and a limited supply of platinum, not more than three dishes were prepared at a time. Platinum electrodes were cleaned and placed in a 91\% alcohol solution between uses. In some instances a control experiment (Fig.~\ref{dish.init}B) was made by connecting the two electrodes of a dish to each other. For all controls, similar zigzag patterns grew above both electrodes.

A potential of 5~V was found to be too strong and introduce mechanical and chemical defects into the experiment. Fig.~\ref{dish.init}C gives an example of the system at 5~V. In this case no E-coli growth appeared above either electrodes. On the other hand, for 1.5~V no discernible effect on the zigzag patterns was found, and they appeared similar to the control experiments. Fig.~\ref{dish.init}D gives an example of an experiment at 1.5~V. 

For all experiments at 3~V, it was found that only one zigzag pattern grew completely, and that it grew selectively around the lower potential electrode. Fig.~\ref{dish.init}E gives an example of a dish after 24~hr at 3~V. We see that the zigzag pattern appears complete only above one electrode. This electrode was connected to the low-potential pole of the power supply. Above the electrode connected to the high-potential pole we find the zigzag pattern is broken up exactly above the electrodes. These results indicate that EM fields play a critical role in the process of biological cell division. Studied of charge movement through the cell membrane during cell division~\cite{hepler1989membranes,strahl2010membrane} point in a similar direction, as do three decades of use of EM for bone fractures~\cite{bassett1991long,griffin2008role}, and more recent studies of EM and Cancer.~\cite{kirson2004disruption}

%One should note that an electric field forms when there is a difference in the EM potential between different regions of space, as was in this experiment. One should also note that this effect is experienced by the ions in the growth medium. Positive ions will accumulate selectively next to the lower potential electrode, and will be selectively absent from around the higher potential electrodes. Since it is known that cells use positive ions to divide~\cite{hepler1989membranes}, in the region close to the higher potential electrode where they are missing, it appears E-coli cells cannot divide.

It is interesting to estimate the amount of charge carriers involved in the division of cells. This is done using previous theoretical results summarized above.~\cite{faraggi2012symmetrical} The distance between the electrodes was 5~cm, hence, for a potential of 3~V the electric field is 60~V/m. It is further assumed that the force on a charge in this field balances the repulsive EM force responsible for cell division. If an estimate for the initial size of the cell is taken as 2~$\mu$m, then the electric field due to a proton in water at this distance is approximately 5~V/m. This indicates that even in a pure dielectric system, relatively little charge of about 12 protons is required to achieve cell division. Accounting for the ionic solution, from Fig.~3 of previous work~\cite{faraggi2012symmetrical} it can be surmised that as little as 2 or 3 net proton charge, symmetrically situated within the cell, is enough to divide it. %However, it is worth mentioning that as a preparation for cell division ionic pumps at the cell surface and the mitotic spindle are responsible for transporting charge to the two dividing daughter cells. It could be argued that a smaller field may be required to interfere with the success of this operation. This is left for future work.

We can estimate the statistical significance of these results by the chi-squared test. Our test statistic in this case is defined as the growth or no-growth of a complete E-coli zig-zag patterns above the electrodes. For 50 dishes with positive results, the p-value is less than 0.0005. This indicates that the results are highly significant and should be easily repeated in other labs. 

To conclude, experiments were performed to test the effect of an electric field on growing E-coli. Petri dishes were fitted with platinum electrodes to create a potential drop across the growing medium. The growth of E-coli colonies was recorded under 1.5, 3, and 5~V. For the 3~V case it was found that E-coli grew preferentially in the vicinity of the lower potential electrode, while there was a clear non-growth area near the higher potential. These results support previous theoretical results indicating that due to charge symmetry in the dividing biological cell electrostatic repulsion is the primary long range repulsive force responsible for the division.~\cite{faraggi2010electrostatic,faraggi2012symmetrical} The results also indicate that electromagnetic fields are crucial in cell division. Continuing to study the relationship between EM and cell division and implementing these results will revolutionize the fields of medicine and biology.

\begin{acknowledgments}
The author would like to gratefully acknowledge informative discussions with Mark Hallet about growing E-coli.
\end{acknowledgments}

%\bibliographystyle{unsrt}
%\bibliography{/home/tmp1/bin/seconds}
%\bibliography{rapidcom}

\begin{thebibliography}{10}

\bibitem{preston1990increased}
Susan Preston-Martin, Malcolm~C Pike, Ronald~K Ross, Peter~A Jones, and Brian~E
  Henderson.
\newblock Increased cell division as a cause of human cancer.
\newblock {\em Cancer research}, 50(23):7415--7421, 1990.

\bibitem{lutkenhaus1997bacterial}
Joe Lutkenhaus and SG~Addinall.
\newblock Bacterial cell division and the z ring.
\newblock {\em Annual review of biochemistry}, 66(1):93--116, 1997.

\bibitem{rothfield1999bacterial}
L~Rothfield, S~Justice, and J~Garcia-Lara.
\newblock Bacterial cell division.
\newblock {\em Annual review of genetics}, 33(1):423--448, 1999.

\bibitem{croce2005mirnas}
Carlo~M Croce and George~A Calin.
\newblock mirnas, cancer, and stem cell division.
\newblock {\em Cell}, 122(1):6--7, 2005.

\bibitem{faraggi2010electrostatic}
Eshel Faraggi.
\newblock An electrostatic model for biological cell division.
\newblock {\em arXiv preprint arXiv:1006.3961}, 2010.

\bibitem{ryaby1978modification}
John~P Ryaby and Arthur~A Pilla.
\newblock Modification of the growth repair and maintenance behavior of living
  tissue and cells by a specific and selective change in electrical
  environment, August~8 1978.
\newblock US Patent 4,105,017.

\bibitem{luben1991effects}
Richard~A Luben.
\newblock Effects of low-energy electromagnetic fields (pulsed and dc) on
  membrane signal transduction processes in biological systems.
\newblock {\em Health Physics}, 61(1):15--28, 1991.

\bibitem{honig1995classical}
Barry Honig, Anthony Nicholls, et~al.
\newblock Classical electrostatics in biology and chemistry.
\newblock {\em Science}, pages 1144--1149, 1995.

\bibitem{kirson2004disruption}
Eilon~D Kirson, Zoya Gurvich, Rosa Schneiderman, Erez Dekel, Aviran Itzhaki,
  Yoram Wasserman, Rachel Schatzberger, and Yoram Palti.
\newblock Disruption of cancer cell replication by alternating electric fields.
\newblock {\em Cancer research}, 64(9):3288--3295, 2004.

\bibitem{bassett1991long}
C~Andrew~L Bassett and Mary Schink-Ascani.
\newblock Long-term pulsed electromagnetic field (pemf) results in congenital
  pseudarthrosis.
\newblock {\em Calcified tissue international}, 49(3):216--220, 1991.

\bibitem{griffin2008role}
Xavier~L Griffin, F~Warner, and M~Costa.
\newblock The role of electromagnetic stimulation in the management of
  established non-union of long bone fractures: what is the evidence?
\newblock {\em Injury}, 39(4):419--429, 2008.

\bibitem{ogura1983partition}
Teru Ogura and Sota Hiraga.
\newblock Partition mechanism of f plasmid: two plasmid gene-encoded products
  and a cis-acting region are involved in partition.
\newblock {\em Cell}, 32(2):351--360, 1983.

\bibitem{lim2005bacterial}
Grace~E Lim, Alan~I Derman, and Joe Pogliano.
\newblock Bacterial dna segregation by dynamic sopa polymers.
\newblock {\em Proceedings of the National Academy of Sciences of the United
  States of America}, 102(49):17658--17663, 2005.

\bibitem{ebersbach2001double}
Gitte Ebersbach and Kenn Gerdes.
\newblock The double par locus of virulence factor pb171: Dna segregation is
  correlated with oscillation of para.
\newblock {\em Proceedings of the National Academy of Sciences},
  98(26):15078--15083, 2001.

\bibitem{barilla2005bacterial}
Daniela Barill{\`a}, Mark~F Rosenberg, Ulf Nobbmann, and Finbarr Hayes.
\newblock Bacterial dna segregation dynamics mediated by the polymerizing
  protein parf.
\newblock {\em The EMBO journal}, 24(7):1453--1464, 2005.

\bibitem{fogel2006dynamic}
Michael~A Fogel and Matthew~K Waldor.
\newblock A dynamic, mitotic-like mechanism for bacterial chromosome
  segregation.
\newblock {\em Genes \& development}, 20(23):3269--3282, 2006.

\bibitem{pratto2008streptococcus}
Florencia Pratto, Aslan Cicek, Wilhelm~A Weihofen, Rudi Lurz, Wolfram Saenger,
  and Juan~C Alonso.
\newblock Streptococcus pyogenes psm19035 requires dynamic assembly of
  atp-bound para and parb on pars dna during plasmid segregation.
\newblock {\em Nucleic acids research}, 36(11):3676--3689, 2008.

\bibitem{ptacin2010spindle}
Jerod~L Ptacin, Steven~F Lee, Ethan~C Garner, Esteban Toro, Michael Eckart,
  Luis~R Comolli, WE~Moerner, and Lucy Shapiro.
\newblock A spindle-like apparatus guides bacterial chromosome segregation.
\newblock {\em Nature cell biology}, 12(8):791, 2010.

\bibitem{hwang2013mediated}
Ling~Chin Hwang, Anthony~G Vecchiarelli, Yong-Woon Han, Michiyo Mizuuchi,
  Yoshie Harada, Barbara~E Funnell, and Kiyoshi Mizuuchi.
\newblock Para-mediated plasmid partition driven by protein pattern
  self-organization.
\newblock {\em The EMBO journal}, 32(9):1238--1249, 2013.

\bibitem{vecchiarelli2013cell}
Anthony~G Vecchiarelli, Ling~Chin Hwang, and Kiyoshi Mizuuchi.
\newblock Cell-free study of f plasmid partition provides evidence for cargo
  transport by a diffusion-ratchet mechanism.
\newblock {\em Proceedings of the National Academy of Sciences},
  110(15):E1390--E1397, 2013.

\bibitem{errington2003cytokinesis}
Jeffery Errington, Richard~A Daniel, and Dirk-Jan Scheffers.
\newblock Cytokinesis in bacteria.
\newblock {\em Microbiology and Molecular Biology Reviews}, 67(1):52--65, 2003.

\bibitem{adams2009bacterial}
David~W Adams and Jeff Errington.
\newblock Bacterial cell division: assembly, maintenance and disassembly of the
  z ring.
\newblock {\em Nature Reviews Microbiology}, 7(9):642--653, 2009.

\bibitem{erickson2010ftsz}
Harold~P Erickson, David~E Anderson, and Masaki Osawa.
\newblock Ftsz in bacterial cytokinesis: cytoskeleton and force generator all
  in one.
\newblock {\em Microbiology and Molecular Biology Reviews}, 74(4):504--528,
  2010.

\bibitem{de2010advances}
Piet~AJ De~Boer.
\newblock Advances in understanding e. coli cell fission.
\newblock {\em Current opinion in microbiology}, 13(6):730--737, 2010.

\bibitem{lutkenhaus2012bacterial}
Joe Lutkenhaus, Sebastien Pichoff, and Shishen Du.
\newblock Bacterial cytokinesis: from z ring to divisome.
\newblock {\em Cytoskeleton}, 69(10):778--790, 2012.

\bibitem{egan2013physiology}
Alexander~JF Egan and Waldemar Vollmer.
\newblock The physiology of bacterial cell division.
\newblock {\em Annals of the New York Academy of Sciences}, 1277(1):8--28,
  2013.

\bibitem{leonard2005towards}
Thomas~A Leonard, Jakob M{\o}ller-Jensen, and Jan L{\"o}we.
\newblock Towards understanding the molecular basis of bacterial dna
  segregation.
\newblock {\em Philosophical Transactions of the Royal Society of London B:
  Biological Sciences}, 360(1455):523--535, 2005.

\bibitem{milo2009bionumbers}
Ron Milo, Paul Jorgensen, Uri Moran, Griffin Weber, and Michael Springer.
\newblock Bionumbers` the database of key numbers in molecular and cell
  biology.
\newblock {\em Nucleic acids research}, 38(suppl\_1):D750--D753, 2010.

\bibitem{faraggi2012symmetrical}
Eshel Faraggi.
\newblock Symmetrical charge-charge interactions in ionic solutions:
  implications for biological interactions.
\newblock {\em arXiv preprint arXiv:1201.0556}, 2012.

\bibitem{jackson}
{J. D. Jackson}.
\newblock {\em {Classical Electrodynamics}}.
\newblock John Wiley and Sons, second edition, 1975.
\newblock pp. 497.

\bibitem{hepler1989membranes}
Peter~K. Hepler.
\newblock {Membranes in the Mitotic Apparatus}.
\newblock In {J. S. Hyams and B. R. Brinkley}, editor, {\em {MITOSIS: Molecules
  and Mechanisms}}, chapter~7, pages 241--271. Academic Press Inc., 1989.

\bibitem{strahl2010membrane}
Henrik Strahl and Leendert~W Hamoen.
\newblock Membrane potential is important for bacterial cell division.
\newblock {\em Proceedings of the National Academy of Sciences},
  107(27):12281--12286, 2010.

\end{thebibliography}

\end{document}